\documentclass[10pt, conference, compsocconf]{IEEEtran}
\usepackage{multirow}

\usepackage{caption}
\renewcommand{\thetable}{\arabic{table}}

\usepackage{cite}
\usepackage{color}
\ifCLASSINFOpdf
  \usepackage[pdftex]{graphicx}
\else
   \usepackage[dvips]{graphicx}
\fi
\usepackage{amssymb,amsmath}
\usepackage{array}
\usepackage{ulem}
\usepackage{url} % hyperref works too
\begin{document}
%\title{Computational Model for Long Bone Growth Initialization}
%\title{Turing pattern formation on a growing deforming domain as a model for long bone development}
\title{Simulating Tissue Morphogenesis and Signaling}


\author{\IEEEauthorblockN{Dagmar Iber, Simon Tanaka, Patrick Fried, Philipp Germann, Denis Menshykau}
\IEEEauthorblockA{Department for Biosystems Science\\ and Engineering (D-BSSE)\\
ETH Zurich\\
Basel, Switzerland\\
E-mail: dagmar.iber@bsse.ethz.ch}
}

\maketitle

%\textcolor{red}{first i would go for the concepts only -  and shift all numerical details to the numerics section. i propose the following sectioning:}

%\tableofcontents

\begin{abstract}
During embryonic development tissue morphogenesis and signaling are tightly coupled. It is therefore important to simulate both tissue morphogenesis and signaling simultaneously in \textit{in silico} models of developmental processes.
The resolution of the processes depends on the questions of interest.
As part of this chapter we will introduce different descriptions of tissue morphogenesis.
In the most simple approximation tissue is a continuous domain and tissue expansion is described according to a pre-defined function of time (and possibly space).
In a slightly more advanced version the expansion speed and direction of the tissue may depend on a signaling variable that evolves on the domain. Both versions will be referred to as 'prescribed growth'. Alternatively tissue can be regarded as incompressible fluid and can be described with Navier-Stokes equations. Local cell expansion, proliferation, and death are then incorporated by a source term. In other applications the cell boundaries may be important and cell-based models must be introduced. Finally, cells may move within the tissue, a process best described by agent-based models. 
\end{abstract}


\IEEEpeerreviewmaketitle

%################################################################################################################################################################################################
\section{Introduction}
%################################################################################################################################################################################################

\begin{figure}[t!]
\begin{centering}
\includegraphics[width=0.8\columnwidth]{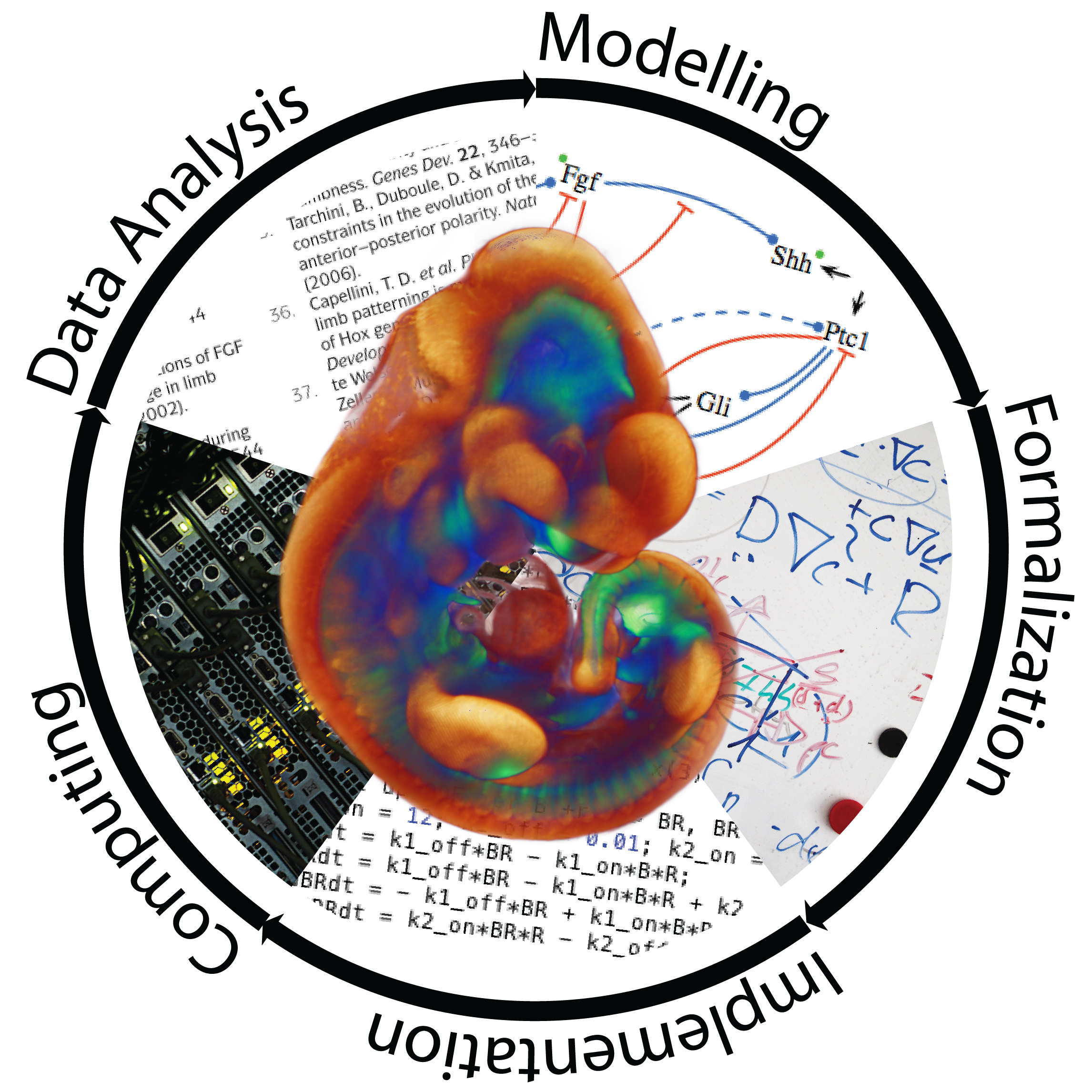}
\par\end{centering}
\caption{\label{fig:Modelling} {\bf \textit{In silico} Models of Tissue Morphogenesis and Signaling.} Models are formulated based on available data. The formalized models then need to be implemented and solved. Model solutions are subsequently compared to available and newly generated data. Models are updated until a good match is achieved. }
\end{figure}

During biological development signaling patterns evolve on dynamically deforming and growing domains. The tissue dynamics affect signaling by advective transport, molecular dilution, separation of signaling centers, and because of the cellular responses to mechanical stress and others. Tissue properties and cellular behaviour, such as cell division and differentiation, in turn are all controlled by the signaling system. To understand the control of tissue growth and organ development both aspects, signaling and tissue mechanics, need to be analysed simultanously. Computational modelling and experimentation are increasingly combined (Figure \ref{fig:Modelling}) to achieve an integrative understanding of such complex processes \cite{Iber:2012hm}. 

Modeling the mechano-chemical interactions mathematically leads to systems, whose numerical solution is challenging. In this review, we present general methods to formulate, couple and solve morphogenetic models. The chapter is organized as follows: In section \ref{sec:signaling} we describe how signaling networks can be modeled on growing and deforming domains using a continuous, deterministic approach. In section \ref{sec:tissuemodels}, different tissue models will be introduced and applications and limitations will be highlighted.

%################################################################################################################################################################################################
\section{Signaling Models on Moving Domains}\label{sec:signaling}
%################################################################################################################################################################################################

Growth can have a significant impact on patterning processes as the growing tissue transports signaling molecules, and molecules are diluted in a growing tissue.
In the following we will discuss the impact of growth on the spatio-temporal distribution of signaling factors.
Let $c_{i}(\boldsymbol{x},t)$ denote the spatio-temporal concentration of a component $i=1,\dots, N$, that can diffuse and react in a volume $\Omega$;
$\boldsymbol{x}$ is the spatial location, and $t$ the time.
The total temporal change of $c_{i}(\boldsymbol{x},t)$ in the volume $\Omega$ must then be equal to the combined changes in the domain due to diffusion and reactions, i.e. 
\begin{equation}
\frac{d}{dt}\int_{\Omega}c_{i}(\boldsymbol{x},t)dV=\int_{\Omega} \{-\nabla\cdot \boldsymbol{j}+R(c_{k}) \} dV 
\label{eq:conservation}
\end{equation}
where $\boldsymbol{j}$ denotes the diffusion flux and $R(c_{k})$ the reaction term, which may depend on the components $c_{k}$, $k=1,\dots,N$.
The molecule $c_{i}$ will diffuse from regions of higher concentration to regions of lower concentration, and we thus have according to Fick's law
\begin{equation*}
\boldsymbol{j}=-D_{i}\nabla c_{i}(\boldsymbol{x},t)
\end{equation*}
which, in case of a constant domain $\Omega$, leads to the well-known reaction-diffusion equation, i.e.
\begin{eqnarray}\label{Eq_diff_constant}
\int_{\Omega} \left \{ \frac{dc_{i}}{dt} - D_{i}\Delta c_{i}-R(c_{k}) \right\} dV = 0 \nonumber \\
\frac{\partial c_{i}}{\partial t} = D_{i}\Delta c_{i}+R(c_{k}).
\end{eqnarray}

If the domain is evolving in time, then the Leibniz integral rule cannot be directly applied. We therefore map the time-evolving domain  $\Omega_{t}$ to a stationary domain $\Omega_{\boldsymbol{\xi}}$ using a time-dependent mapping. $\boldsymbol{\xi}$ denotes the spatial coordinate in the statinonary domain. For the left hand side of eq. \eqref{eq:conservation} we then obtain, using the Reynolds transport theorem,

\begin{eqnarray*}
\frac{d}{dt}\int_{\Omega_{t}}c_{i}(\boldsymbol{x},t)\: d\Omega & = & \frac{d}{dt}\int_{\Omega_{\boldsymbol{\xi}}}c_{i} \left( \boldsymbol{x}(\boldsymbol{\xi},t),t\right) J\: d\Omega\\
 & = & \int_{\Omega_{\boldsymbol{\xi}}}\left[\frac{dc_{i}}{dt}J+c_{i}\frac{dJ}{dt}\right]d\Omega\\
 & = & \int_{\Omega_{\boldsymbol{\xi}}}\left[\frac{\partial c_{i}}{\partial t}+\boldsymbol{u} \cdot \nabla c_{i}+c_{i} \nabla\cdot \boldsymbol{u}\right]J\: d\Omega\\
 & = & \int_{\Omega_{t}}\left[\frac{\partial c_{i}}{\partial t}+\nabla \cdot (c_{i}\boldsymbol{u})\right]d\Omega
\end{eqnarray*}
\noindent where $J$ with $\dot{J}=J\nabla \boldsymbol{u}$ denotes the Jacobian and $\boldsymbol{u} = \frac{\partial \boldsymbol{x}}{\partial t}$ the velocity field.
We thus obtain as reaction-diffusion equation on a growing domain:
\begin{equation}\label{eq:eulerian}
\left . \frac{\partial c_{i}}{\partial t}\right |_{\boldsymbol{x}}+\nabla\cdot(c_i\boldsymbol{u})=D_{i}\Delta c_{i} + R(c_{i}).
\end{equation}
$|_{\boldsymbol{x}}$ indicates that the time derivative is performed while keeping $\boldsymbol{x}$ constant.
The terms $\boldsymbol{u}\cdot\nabla c_{i}$ and $c_{i}\nabla\cdot\boldsymbol{u}$ describe advection and dilution, respectively.
If the domain is incompressible,  i.e. $\nabla\cdot\boldsymbol{u}=0$, the equations further simplify. 

It should be noted that this deterministic reaction-diffusion equation only describes the mean trajectory of an ensemble.
Whenever the molecular population of the least prevalent compound is small, the advection-diffusion equation is not a good description and stochastic techniques need to be used.\\

%--------------------------------------------------------------------------------------------------------------------------------------

\subsection{The Lagrangian Framework}

In growing tissues cells move. It can be beneficial to take the point of view of the cells and follow them. This is possible within the Lagrangian Framework. To illustrate the differences between the Eulerian and Lagrangian framework consider a river. The Eulerian framework would correspond to sitting on a bench and watching the river flow by. In the Lagrangian framework we would sit in a boat and travel with the river. 

Accordingly, at time $t=0$ we now label a particle by the position vector $\boldsymbol{X}= \boldsymbol{x}(0)$ and follow this particle over time. At times $t>0$, the particle is found at position $\boldsymbol{x} = \psi \left(\boldsymbol{X},t\right)$. Here $\boldsymbol{x}$ is the spatial variable in the Eulerian framework and $\boldsymbol{X}$ is the spatial variable in the Lagrangian framework. If initially distinct points remain distinct throughout the entire motion then the transformation possesses the inverse $\boldsymbol{X} = \psi^{-1} \left(\boldsymbol{x},t\right)$. Any quantity $F$ (i.e. a concentration $F=c_i$) can therefore be written either as a function of Eulerian variables $(\boldsymbol{x},t)$ or Lagrangian variables $(\boldsymbol{X},t)$. To indicate a particular set of variables we thus write either $F=F(\boldsymbol{x}(\boldsymbol{X},t),t)$ as the value of $F$ felt by the particle instantaneously at the position $\boldsymbol{x}$ in the Eulerian framework, or  $F=F(\boldsymbol{X},t)$ as the value of $F$ experienced at time $t$ by the particle initially at $\boldsymbol{X}$ (Lagrangian Framework).

In the Lagrangian framework we now need to determine the change of the variable $F$ following the particle, while in the Eulerian framework we were determining $\left .\frac{\partial F}{\partial t} \right|_{\boldsymbol{x}}$, the rate of $F$ apparent to a viewer stationed at the position $\boldsymbol{x}$.  The time derivative in the Lagrangian framework is also called the material derivative:
\begin{equation}\label{eq:lagrangian_material_derivative}
 \frac{dF}{d t} = \frac{dF(\boldsymbol{x}(\boldsymbol{X},t),t)}{d t} =\frac{\partial F(\boldsymbol{X},t)}{\partial t}
\end{equation}
and follows as
\begin{eqnarray}\label{eq:lagrangian_time_derivative}
\underbrace{ \frac{dF(\boldsymbol{X},t)}{d t}}_{Lagrangian}  &=&
\left .\frac{\partial F}{\partial t} \right|_{\boldsymbol{x}} +\frac{\partial F}{\partial x_k} \underbrace{\frac{\partial x_k(\boldsymbol{X},t)}{\partial t}}_{=u_k} \nonumber \\
&=& \underbrace{\left .\frac{\partial F}{\partial t} \right|_{\boldsymbol{x}} + \boldsymbol{u} \cdot \nabla F}_{Eulerian}. 
\end{eqnarray}
Note that the advection term $\boldsymbol{u}\cdot\nabla F$ vanishes in the material derivative as compared to the Eulerian description. We can now also write the Eulerian spatial derivatives in terms of the Lagrangian reference frame using the Jacobian of the transformation
\begin{equation}
J = \frac{\partial(X_1, X_2, X_3)}{\partial(x_1, x_2, x_3)}.
\end{equation}
Geometrically, $J$ represents the dilation of an infinitesimal volume as it follows the motion:
\begin{equation}
d X_1 dX_2 dX_3 = J dx_1 dx_2 dx_3. 
\end{equation}

\begin{figure}[t!]
\begin{centering}
\includegraphics[width=0.6\columnwidth]{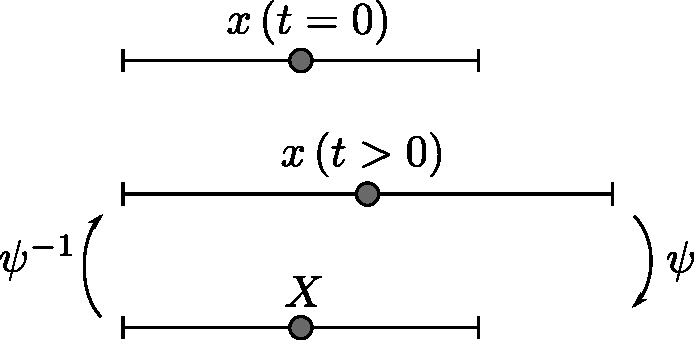}
\par\end{centering}
\caption{ {\bf Mapping to a Stationary Domain.}
A one dimensional domain is stretched. A point on the domain, initially at $x\left(t=0\right)$ is advected and later found at position $x\left(t>0\right)$.
At all times, the Eulerian coordinate system can be mapped to a stationary domain using a mapping function $\psi$, and vice versa using its inverse $\psi^{-1}$.
On the stationary domain, the point stays at the same position for all times and thus can be labeled by $X$.
}
\label{fig:stationarydomainmapping}
\end{figure}

\paragraph{Example - Uniform Growth}
The benefit of working in a Lagrangian reference frame is directly apparent in case of a uniformly growing domain.
In case of uniform growth in one spatial dimension we have $x=L(t)X$, where $L(t)$ is the time-dependent length of the domain. We then have
\begin{equation}
\frac{\partial X}{\partial x}=\frac{1}{L(t)} \hspace{1cm}
u=\dot{L(t)}X \hspace{1cm} \frac{\partial u}{\partial X}=\dot{L(t)}
\end{equation}
Since the stretching factor $L(t)$ is independent of the spatial position, the Lagrangian reference frame $X$ corresponds to a stationary domain. As reaction-diffusion equation on an uniformly growing domain we then obtain a rather simple formula, i.e.
\begin{equation}
\frac{dc}{dt} + \frac{\dot{L(t)}}{L(t)}c = D \frac{1}{L(t)^{2}}\frac{\partial^{2}c}{\partial X^{2}}+R(c)
\end{equation}
where $c = c(X,t)$. The principle is summarized in Figure \ref{fig:stationarydomainmapping}.
We have used this approach in a 1D model of bovine ovarian follicle development (Iber and De Geyter, under review).\\

\subsection{Arbitrary Lagrangian-Eulerian (ALE) Method}
The arbitrary Lagrangian-Eulerian (ALE) method is a generalization of the well-known Eulerian and Lagrangian domain formulations  \cite{Donea2004}.
In the Eulerian framework, the observer does not move with respect to a reference frame (Equation \ref{eq:eulerian}).
Large deformations can be described in a simple and robust way, but tracking moving boundaries can lead to non-trivial problems.
In the Lagrangian framework, on the other hand, the observer moves according to the local velocity field.
The convective terms are zero because the relative motion to the material vanishes locally, and the equations simplify substantially (Equation \ref{eq:lagrangian_time_derivative}). However, this comes at the expense of mesh distortions when facing large material deformations.

\begin{figure}
\begin{centering}
\includegraphics[width=0.4\columnwidth]{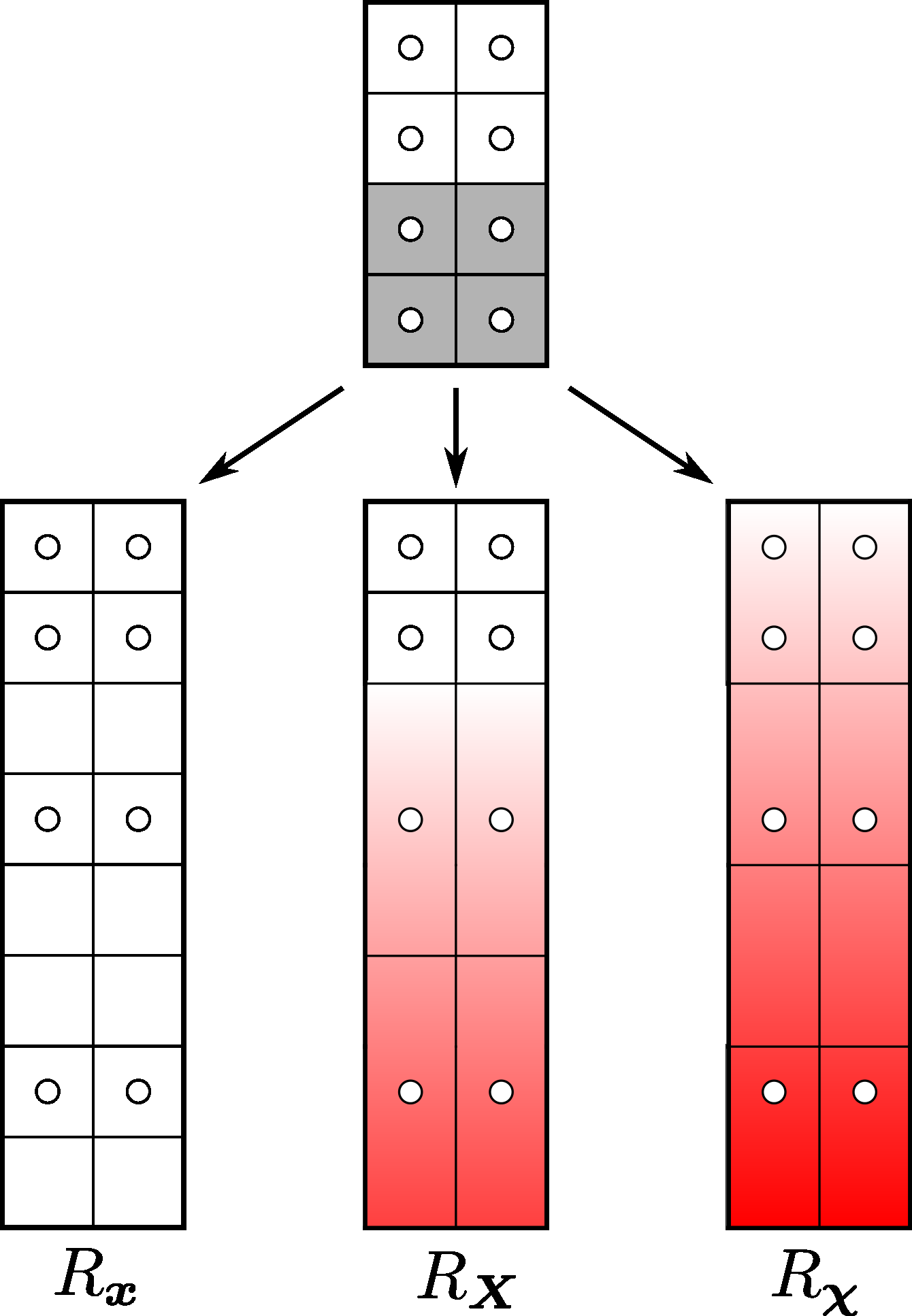}
\par\end{centering}
\caption{ {\bf Reference Frame Paradigms.}
The grey shaded material of the initial domain is stretched threefold. Material particles (circles) are attached to the continuum. In the Eulerian domain $R_{\boldsymbol{x}}$ the mesh does not move as opposed to the Lagrangian domain $R_{\boldsymbol{X}}$ and ALE domain $R_{\boldsymbol{\chi}}$. The red color denotes the magnitude of mesh velocity $\boldsymbol{v}$. In the Lagrangian domain, the mesh velocity coincides with the material velocity field $\boldsymbol{u}$,
whereas in the ALE domain the mesh velocity can be chosen arbitrarily.}
\label{fig:ALE_principle}
\end{figure}

In the ALE framework, finally, the observer is allowed to move freely and describe the equations of motions from his viewpoint.
This allows for the flexibility to deform the mesh according to e.g. moving boundaries, but also for the possibility to freely remodel the mesh independent of the material deformations.
Although the problem of mesh distortion is much reduced as compared to the Lagrangian formulation, remeshing might still be required when confronted with complex deformations.
The three paradigms are visualized in Figure \ref{fig:ALE_principle}.

In the ALE framework, the reaction-diffusion equation reads:
\begin{equation}
\left. \frac{\partial c_{i}}{\partial t} \right|_{\boldsymbol{x}} + \boldsymbol{w} \cdot \nabla c_{i} + c_{i} \nabla\cdot\boldsymbol{u} = D_{i} \Delta c_{i} + R\left(c_{i}\right)
\end{equation}

\noindent where $\left. \partial_{t} c_{i} \right|_{\boldsymbol{x}}$ denotes the time derivative with fixed $\boldsymbol{x}$ coordinate. $\boldsymbol{w} = \boldsymbol{u}-\boldsymbol{v}$ is the convective velocity (i.e. the relative velocity between the material and the ALE frame) and $\boldsymbol{v}$ the mesh velocity.
In the case of $\boldsymbol{v} \equiv \boldsymbol{u}$, i.e. the mesh is attached to the material, the Lagrangian formulation (Equation \ref{eq:lagrangian_time_derivative}) is recovered.
On the other hand, when setting $\boldsymbol{v} \equiv 0$, we get back the Eulerian formulation (Equation \ref{eq:eulerian}). In between, the mesh velocity $\boldsymbol{v}$ can be chosen freely, which can be exploited to being able to track large deformations.\\

\begin{figure*}[t!]
\begin{centering}
\includegraphics[width=1\textwidth]{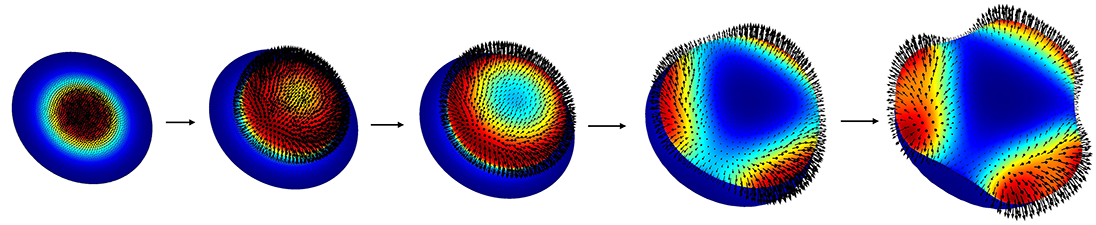}
\par\end{centering}
\caption{\label{fig:growth} {\bf 'Prescribed' Domain Growth under Control of a Signaling Model.} The deformation of the domain is controlled by a Turing-type signaling model (Equation \ref{eq:schnak}) according to $\boldsymbol{u}= \mu c_1^2c_2 \boldsymbol{n}$. The red and blue regions denote areas with high and low concentration of $c_1^2 c_2$; the arrows denote the velocity field.}
\end{figure*}

%################################################################################################################################################################################################
\section{Tissue Models}\label{sec:tissuemodels}
%################################################################################################################################################################################################

%=================================================================================
\subsection{Prescribed Growth}
%=================================================================================

The development of mechanistic models of tissue growth is challenging and requires detailed knowledge of the gene regulatory network, mechanical properties of the tissue, and its response to physical and biochemical cues.
If these are not available but the expansion of the tissue has been described, a phenomenological approach can be used to prescribe the geometry based on observations. 

In 'prescribed growth models' an initial domain and a spatio-temporal velocity or displacement field are defined.
The domain with initial coordinate vectors $\boldsymbol{X}$ is then moved according to this velocity field $\boldsymbol{u}(\boldsymbol{X},t)$, i.e. 

\begin{equation}
\frac{ \partial \boldsymbol{X}(t)}{\partial t} = \left. \frac{\partial \boldsymbol{x}}{\partial t} \right|_{\boldsymbol{X}} =  \boldsymbol{u}(\boldsymbol{X}, t)
\end{equation}

%=================================================================================
\subsubsection{Model-based Displacement Fields}
%=================================================================================

The velocity field $\boldsymbol{u}(\boldsymbol{X}, t)$ can be captured in a functional form that represents either the observed growth and/or signaling kinetics.
In the simplest implementation the displacement may be applied only normal to the boundary, i.e. $\boldsymbol{u}= \mu \boldsymbol{n}$, where $\boldsymbol{n}$ is the normal vector to the boundary and $\mu$ is the local growth rate.
We studied such models in the context of organ development and found that the patterning on the developing lung and limb domains depends on the growth speed \cite{Probst:2011jo,Menshykau:2012kg,Celliere:2012jc, Badugu:2012ho}. 

Growth processes often depend on signaling networks that evolve on the tissue domain. The displacement field $\boldsymbol{u}(\boldsymbol{X},t)$ may thus dependent on the local concentration of some growth or signaling factor. We then have $\boldsymbol{u}= \mu(c) \boldsymbol{n}$ where $c$ is the local concentration of the signaling factor.
These approaches can be readily implemented in the commercially available finite element solver COMSOL Multiphysics; details of the implementation are described in \cite{Germann:bT_kMV7D, Menshyau2012c}. Figure \ref{fig:growth} shows as an example a 2D sheet that deforms within a 3D domain according to the strength of the signaling field normal to its surface, i.e. $\boldsymbol{u}= \mu c_1^2c_2 \boldsymbol{n}$, where $c_1$ and $c_2$ are the two variables that are governed by the Schnakenberg-type Turing model 
\begin{eqnarray}\label{eq:schnak} 
\frac{\partial c_{1}}{\partial t} + \nabla\cdot(c_1\boldsymbol{u}) &=& \Delta c_1+\gamma(a-c_1+c_1^2c_2) \nonumber \\
\frac{\partial c_{2}}{\partial t} + \nabla\cdot(c_2\boldsymbol{u}) &=&d \Delta c_2+\gamma(b-c_1^2c_2);
\end{eqnarray}
$a$, $b$, $\gamma$, and $d$ are constant parameters in the Turing model. \\

\begin{figure*}[t!]
\begin{centering}
\includegraphics[width=1\textwidth]{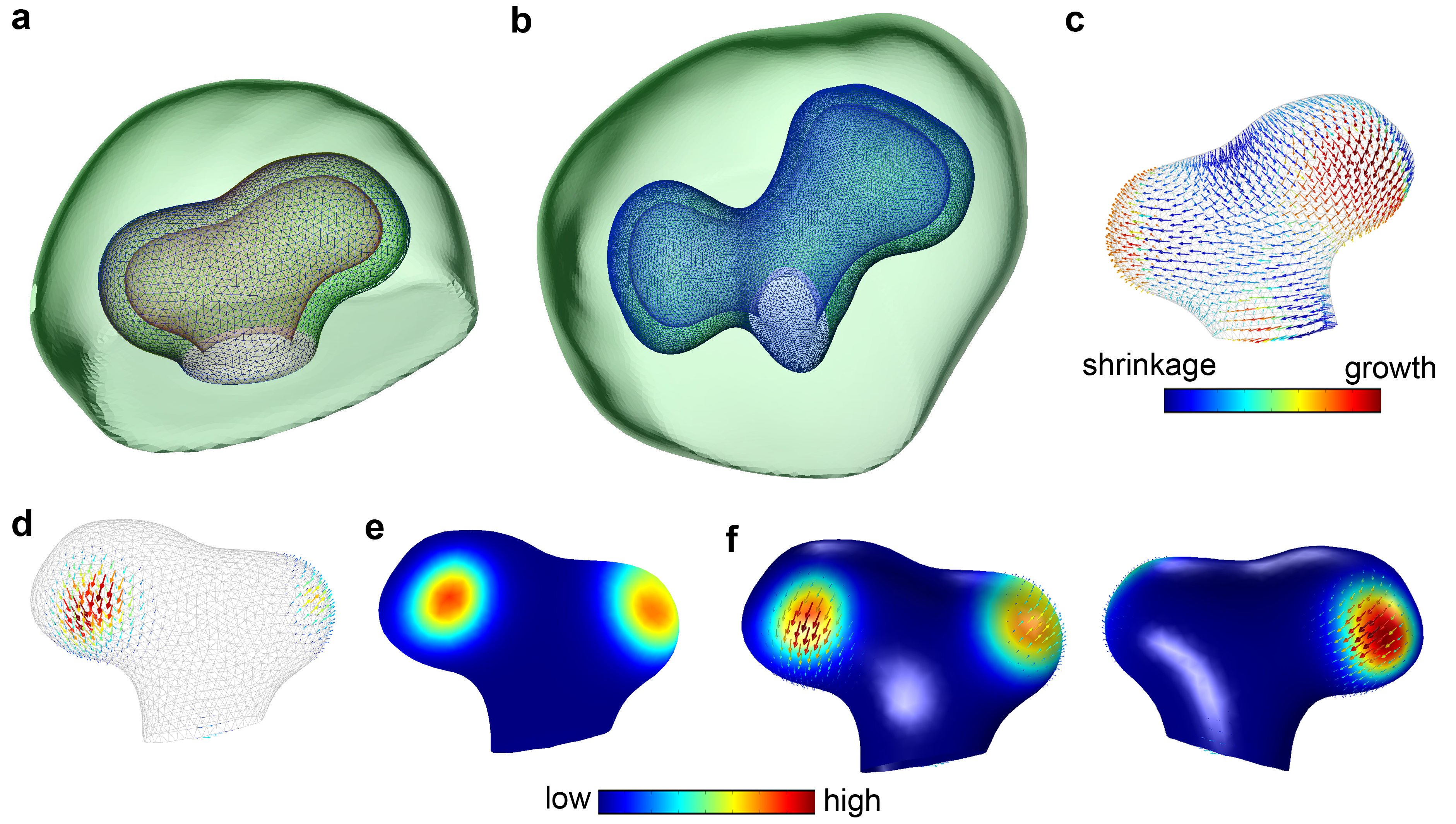}
\par\end{centering}
\caption{\label{fig:image_based} {\bf Image-based Displacement Fields.} (a,b) The segmented epithelium and mesenchyme of the developing lung at two consecutive stages. (c) The displacement field between the two stages in panels a and b. (d) The growing part of the lung. The coloured vectors indicate the strength of the displacement field. (e) The solution of the Turing model (Equations \ref{eq:schnak}) on the segmented lung of the stage in panel a. (f) Comparison of the simulated Turing model (solid surface) and the embryonic displacement field (arrows). The images processing was carried out in AMIRA; the simulations were carried out in COMSOL Multiphysics 4.3a. The panels in the figure have been reproduced from Menshykau et al, submitted.}
\end{figure*}

%=================================================================================
\subsubsection{Image-based Displacement Fields}
%=================================================================================

The displacement field may also be obtained from experimental data. To obtain the displacement field from data, tissue geometries need to be extracted at sequential time points as shown for lung development in Figure \ref{fig:image_based}a,b. This requires the following steps: 1) staining of the tissue of interest, 2) imaging of the tissue at distinct developmental time points, 3) image segmentation, 4) meshing of the segmented domain, 5) warping (morphing) of images at various developmental stages. Subsequently a mathematical regulatory network model can be solved on the deforming physiological domain. In the following we will discuss the different steps in detail.\\

\paragraph{3D Image and Meshes of Tissue}
In the first step we need to obtain 3D imaging data of the tissue of interest. In case different sub-structures are of interest, the tissue needs to be labelled accordingly. The staining and imaging technique of choice depends on the tissue, the sub-structure of interest, and the  desired resolution. Available techniques have been reviewed in depth before \cite{Gregg:2012jy}.

Once the imaging data has been obtained these need to be processes computationally to obtain the 4D datasets. Several image processing software packages are available to perform these steps, e.g. Amira or Imaris. If multiple image recordings of the organ or tissue are available at a given stage, then the 3D images can be aligned and averaged. The alignment procedure is a computationally non-trivial problem. In Amira a number of iterative hierarchical optimization algorithms (e.g. QuasiNewton) are available as well as similarity measures (e.g. Euclidean distance) to be minimized. Averaging is subsequently performed by averaging pixel intensities of corresponding pixels in multiple datasets of the same size and resolution. This helps to assess the variability between embryos and identifies common features. It also reduces variability due to experimental handling, but averaging of badly aligned datasets can result in loss of biologically relevant spatial information. It is therefore suggested to run the alignment algorithm several times, starting with different initial positions of the objects, which are to be aligned. 

The next step is to perform image segmentation. During image segmentation the digital image is partitioned into multiple subdomains, usually corresponding to anatomic features and gene expression regions. A variety of algorithms are available for image segmentation,  most of which are based on differences in pixel intensity. 

To carry out finite element methods (FEM)-based simulations of the signaling networks, segmented images are subsequently converted into meshes of sufficient quality. The quality of the mesh can be assessed according to the following two parameters: mesh size and the ratio of the sides of the mesh elements. The linear size of the mesh should be much smaller than any feature of interest in the computational solution, i.e. if the gradient length scale in the model is 50 $\mu m$ then the linear size of the mesh should be at least several times less than 50 $\mu m$. Additionally, the ratio of the length of the shortest side to the longest side should be 0.1 or more. To confirm the convergence of the simulation, the model must be solved on a series of refined meshes.\\

\paragraph{Calculating the Displacement Field}
To simulate the signaling models on growing domains we need to determine the displacement fields between the different stages. The displacement field between two consecutive stages can be calculated by morphing two subsequent stages onto each other. In other words we are looking for a function which returns a point on a surface at time $t+\Delta t$ which corresponds to a point on a surface at time $t$.

The landmark-based Bookstein algorithm \cite{Bookstein:1989wl}, which is implemented in Amira, uses paired thin-plate splines to interpolate surfaces over landmarks defined on a pair of surfaces. The landmark points need to be placed by hand on the two 3D geometries to identify corresponding points on the pair of surfaces. The exact shape of the computed warped surface therefore depends on the exact position of landmarks; landmarks must therefore be placed with great care. While various stereoscopic visualization technologies are available this process is time-consuming and in parts difficult for complex surfaces such as the epithelium of the embryonic lung or kidney, in particular if the developmental stages are further apart. 

Once the correspondence between two surfaces has been defined, a displacement field can be calculated by determining the difference between the positions of points on the two surface meshes as illustrated for the embryonic lung sequence  in Figure \ref{fig:image_based}c; panel d highlights the growing part of the lung. \\

\paragraph{Simulation of Signaling Dynamics using FEM}
To carry out the FEM-based simulations the mesh and displacement field need to be imported into a FEM solver.
To avoid unnecessary interpolation of the vector field, the displacement field should be calculated for exactly the same surface mesh as was used to generate the volume mesh. A number of commercial (COMSOL Multiphysics, Ansis, Abaqus etc) and open (FreeFEM, DUNE etc) FEM solvers are available. Figure \ref{fig:image_based}e shows the solution of the Schnakenberg Turing model (Equations \ref{eq:schnak}) on the segmented lung of the stage in panel a. The distribution of the simulated Turing pattern coincides with the embryonic displacement field as shown as arrows (Figure \ref{fig:image_based}f).\\

%=================================================================================
\subsection{Continuous Tissue Models}
%=================================================================================

\begin{figure}[t!]
\begin{centering}
\includegraphics[width=1\columnwidth]{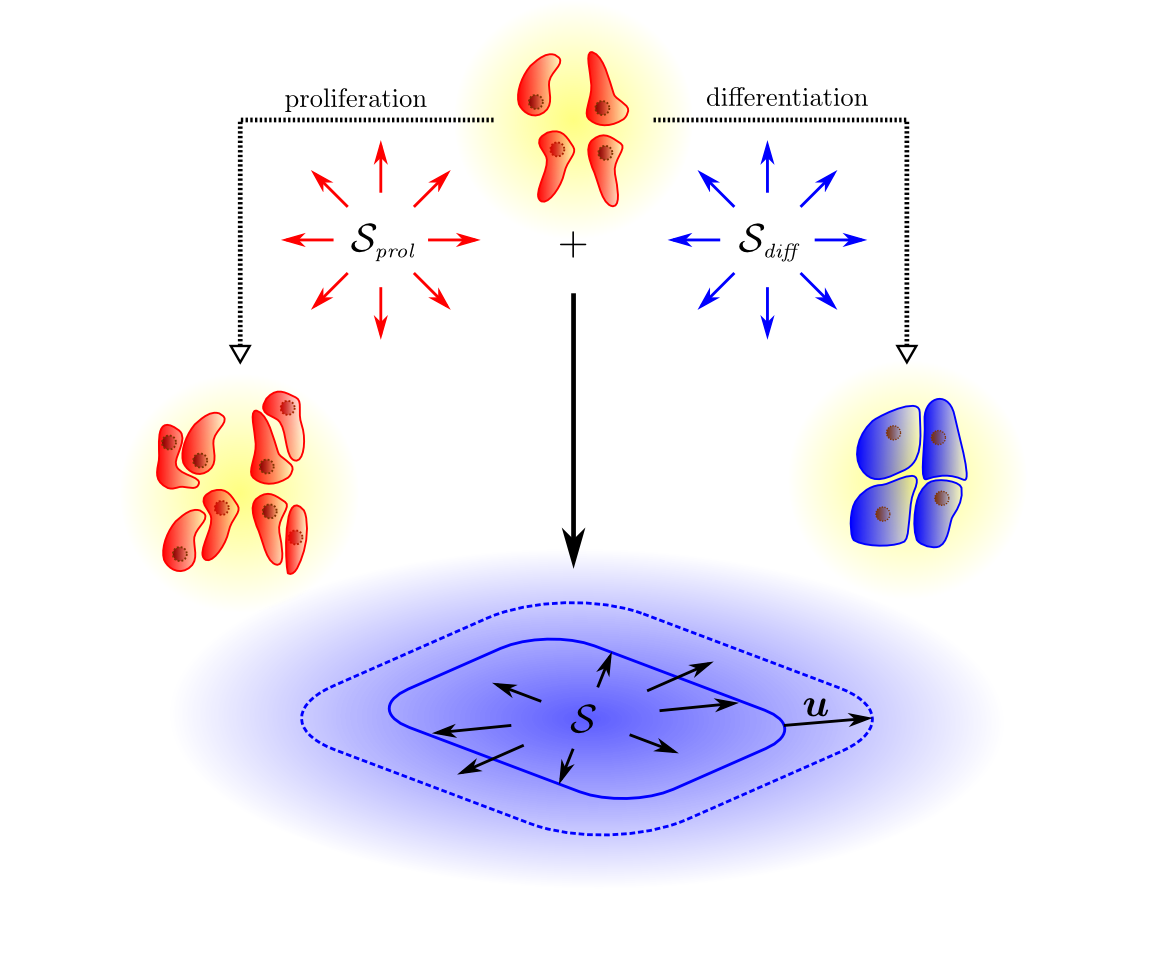}
\par\end{centering}
\caption{{\bf Tissue as an incompressible fluid.}
		Proliferating cells (shown in red) may divide, which is modeled as a local mass source $\mathcal{S}_{\textit{prol}}$ (left path).
		As a result of differentiation, the cells increase in volume and lead to a local mass source $\mathcal{S}_{\textit{diff}}$ (right path).
		Both mechanisms induce a velocity field $\boldsymbol{u}$ in the fluid.}
\label{fig:tissuemodel}
\end{figure}

In an alternative approach tissue is treated as an incompressible fluid with fluid density $\rho$, dynamic viscosity $\mu$, internal pressure $p$, and fluid velocity field $ \boldsymbol{u}$.
Tissue can then be described by the Navier-Stokes equation:

\begin{subequations}
\begin{align}
\rho \left( \partial_{t} \boldsymbol{u} + \left( \nabla \cdot \boldsymbol{u} \right) \boldsymbol{u} \right) &=
		-\nabla p + \mu \left( \Delta \boldsymbol{u} + \frac{1}{3}\nabla\left(\nabla\cdot \boldsymbol{u}\right) \right) + \boldsymbol{f} \label{eq:momentumequation}\\
\rho \nabla \cdot \boldsymbol{u} &= \omega \mathcal{S}
\end{align}
\end{subequations}

\noindent  where $\omega \mathcal{S}$ denotes the local mass production rate, which is composed of contributions from proliferation, $\mathcal{S}_{prol}$, and increase in cell volume by cell differentiation, $\mathcal{S}_{diff}$ (Figure \ref{fig:tissuemodel}). $\omega$ is the molecular mass of cells, $\left [\frac{kg}{mol}\right ]$. The impact of cell signaling on tissue morphogenesis can be implemented via the source term $ \mathcal{S}= \mathcal{S}_{prol}+\mathcal{S}_{diff}$ in that $ \mathcal{S}$ can depend on the local concentration of growth or differentiation factors. $\boldsymbol{f}$ denotes the external force density and may e.g. originate from cellular structures which exert force on the fluid. 

The dynamic viscosity $\mu$ of embryonic tissue is approximately $\mu \approx 10^{4} \left[Pa\cdot s\right]$ \cite{Forgacs:1998fy}, some $10^7$-fold higher than for water, and the mass density of the tissue is $\rho \approx 1000 \left[kg/m^{3} \right]$.
Using a characteristic reference length $L$ and a characteristic reference speed $U$, the non-dimensional Reynolds number $Re=\rho LU/\mu$ is estimated to be of order $10^{-14}$ in typical embryonic tissue.
The Reynolds number characterizes the relative importance of inertial over viscous forces, whereby the latter are dominant in tissue mechanics. After non-dimensionalization, the Navier-Stokes equation (\ref{eq:momentumequation}) reads (for the now non-dimensional variables $\boldsymbol{u}$ and $p$)
\begin{equation}\label{eq:nondimnavierstokes}
 Re \left(\partial_{t} \boldsymbol{u} + \left(\nabla \cdot \boldsymbol{u} \right) \boldsymbol{u}\right) = -\nabla p + \Delta \boldsymbol{u} + \frac{1}{3}\nabla \left(\nabla\cdot \boldsymbol{u}\right).
\end{equation}
Since $Re$ is very small, the left hand side of equation (\ref{eq:nondimnavierstokes}) can be neglected, resulting in the well-known Stokes equation for creeping flow. The Navier-Stokes equations can be numerically solved using finite diffference methods (FDM), finite element methods (FEM), finite volume methods (FVM), spectral methods, particle methods and Lattice-Boltzmann methods (LBM) \cite{Chen1998}. 

The Navier-Stokes description has been used in simulations of early vertebrate limb development \cite{Dillon:2003p12910}, and, in an extended anisotropic formulation, has been applied to Drosophila imaginal disc development \cite{Bittig:2008uc}. In case of the limb the applicability of an isotropic Navier-Stokes model to tissue growth has been challenged by experimental measurements \cite{Boehm:2010p42421}. To that end Boehm and collaborators determined the proliferation rates inside the limb and used the measured rates as source terms in the isotopic Navier-Stokes tissue model. They then compared the predicted shapes to measured shapes and noticed large discrepancies. They subsequently solved the inverse problem to obtain $\mathcal{S}$ from the measured shapes and found that $\mathcal{S}$ needed to also take negative values, and that the expansion was larger than expected from the measured proliferation rates. Limb expansion thus must result from anisotropic processes that also involve cell migration from the flank.

%=================================================================================
\subsection{Cell-Based Tissue Models}
%=================================================================================

All approaches described above neglect that tissues are an ensemble of cells. While many effects that result from cell-cell interactions can be described also with continuous differential equations, cell-based tissue models permit a detailed, mechanistic description of the process that relates more easily to the biophysical measurements, and that can help to understand how observed macroscopic properties may emerge from the microscopic interactions. Such cell based simulations also allow simulations to explore signal read-outs on a discrete cell level where receptors can diffuse on the surface of a cell but not between cells. 

Most cell based models are hybrid models that capture the discrete, individual nature of cells and which also include partial differential equations (PDEs) that give a continuous description of signaling pathways or availability of nutrients.
These models have the advantage that they integrate biological processes happening on different scales, i.e., they describe signaling processes within cells, forces between cells and observe effects on a multicellular level.

There are two general ways of how to define cells: Lattice-based approaches where a cell occupies a certain number of lattice entities, e.g., squares or hexagons and off-lattice approaches where cells can occupy an unconstrained area/volume in the 2D/3D space.\\

\subsubsection{Viscoelastic Cell Model}
\begin{figure}[t!]
\begin{centering}
\includegraphics[width=1\columnwidth]{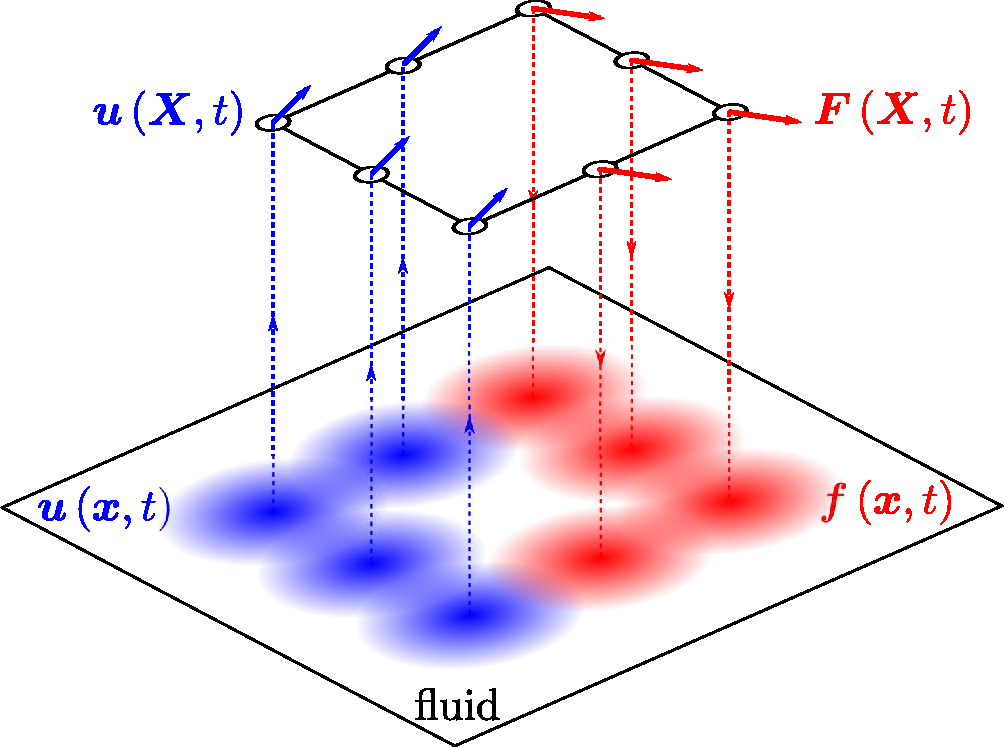}
\par\end{centering}
\caption{{\bf Immersed Boundary Method.}
		The geometry is discretized into nodes at positions $\boldsymbol{X}$.
		The force density $\color{red} \boldsymbol{F}\left(\boldsymbol{x},t\right)$, hosted by the node, is distributed to the local fluid neighborhood using a delta Dirac kernel function.
		The nodes are moved according to the local velocity $\color{blue} \boldsymbol{u}\left(\boldsymbol{X},t\right)$,
		which is computed from the fluid velocity $\color{blue} \boldsymbol{u}\left(\boldsymbol{x},t\right)$ using the same kernel function.
		}
\label{fig:IB_principle}
\end{figure}

Elastic cellular components such as the membrane and cell junctions, play a key role in the cellular dynamics. The core idea of the viscoelastic cell model, introduced in \cite{Dillon2000,Rejniak2004,Rejniak2007}, is to divide the viscous and elastic properties and represent these by a viscous fluid and massless elastic structures, respectively.
The latter are modeled as elastic networks, which exert forces on the fluid.
The fluid, on the other hand, exerts force on the elastic structures, which leads to a classic fluid-structure-interaction (FSI) problem. A well-known technique to solve FSI problems is the immersed boundary (IB) method \cite{Peskin2003}, which is illustrated in Figure \ref{fig:IB_principle}. The boundary is discretized into computational boundary nodes, which spread the force to their local neighborhood defined by a delta Dirac kernel function.
Apart from the forcing term $\boldsymbol{f}$ in Equation (\ref{eq:momentumequation}), the fluid does not 'see' the boundary, which significantly facilitates the numerical solution of the problem.
The boundary nodes are subsequently moved in a Lagrangian manner according to the local velocity field. 

Although the high computational costs limit this approach to intermediate problem sizes (up to few thousand cells) as compared to continuous cell-density representations and models with rudimentary cell representations,
the simulation parameters, e.g. membrane elasticity and interstitial fluid and cytoplasm viscosity, can be inferred directly from biophysical measurements, as opposed to more abstract approaches.
The method has been deployed to study, amongst others, tumor growth and ductal carcinoma development\cite{Dillon2000}, growth of the trophoblast bilayer \cite{Rejniak2004} and formation of epithelial hollow acini \cite{Rejniak2008,Rejniak2008a}.\\

\subsubsection{Cellular Potts Model}
One important example for a lattice-based method is the Monte-Carlo-based Cellular Potts Model (CPM) \cite{Graner1992}, which is implemented in the modeling framework CompuCell3D \cite{Izaguirre2004}. CompuCell3D models both cell behaviour and signaling dynamics by coupling the CPM module to a PDE module for diffusible signaling factors. 

In the CPM framework every cell is represented by a set of lattice sites $\vec{i}$. Cell expansion is represented by an increase of lattice sites per cell. As one cell expands another cell will shrink by one lattice site. If both cell types represent cells in the tissue the overall tissue size stays constant. Tissue growth can be achieved by introducing one cell type that represents the medium and that subsequently loses lattice sites to the cells in the tissue. Cell movement is achieved by a shift of the cell-specific lattice sites (identified by the cell index $\sigma(\vec{i})$) along the lattice. Each cell belongs to a specified cell type with index $\tau(\sigma (\vec{i}))$. Cells can secret, interact with, and respond to the diffusible signaling factors.

\begin{figure}[t!]
\begin{centering}
\includegraphics[width=1\columnwidth]{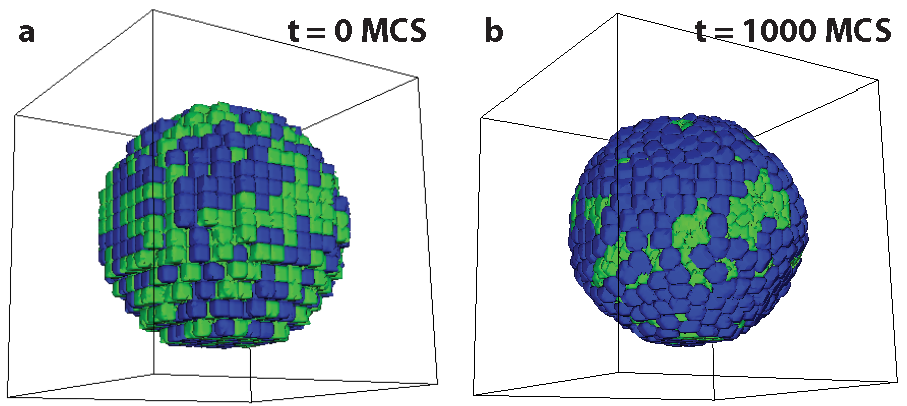}
\par\end{centering}
\caption{{\bf Cellular Potts Model.} 
	A typical CPM simulation of cell sorting using CompuCell3D. (a) Random initial configuration with two different cell types depicted in blue and green. Both cell types share the same properties, but the contact energy between cells of the same cell type is lower compared to the contact energy between mixed cell types. (b) After 1000 MCS cells of a given type have clustered together to minimize the total energy, resulting in a patch of blue cells on one side of the sphere; the total number of cells per cell type is unchanged.}
\label{fig:CPM}
\end{figure}

CompuCell3D implements a variant of the Metropolis Monte Carlo method. In every time step of the model, also called Monte Carlo sweep, on average every lattice site can attempt a transition to a different state. Thus in case of $N$ lattice sites, during each sweep $N$ lattice sites $\vec{i}$ and a neighboring lattice site $\vec{j}$ are chosen at random. If the cell indices $\sigma(\vec{i})$ and $\sigma(\vec{j})$ are different then a new configuration is proposed in which the neighboring lattice site becomes part of the originally chosen cell, i.e. its cell index changes to $\sigma(\vec{j}) = \sigma(\vec{i})$. Every proposed new configuration is accepted with the probability 
\begin{equation}
P=\text{min}(1,\exp(-\frac{\Delta E}{kT})).
\end{equation} 
This means that proposed moves which lower the energy ($\Delta E< 0$) are always accepted, while moves, which increase the energy ($\Delta E>0$) are accepted with a probability that depends on the energy difference $\Delta E$ and the energy scaling factor $kT$. The energy of a configuration includes different energy terms, e.g. adhesion is calculated by the sum of  the contact energies per unit area $J(\tau,\tau ')$, which depends on the cell types that are in contact. In case of  cell types with high adhesive forces, represented by low contact energies, cell clusters will emerge as these minimize the overall contact energy (Figure \ref{fig:CPM}).

The energy scaling factor $kT$ controls how easily energetically unfavorable configurations are accepted. If $kT$ is very large, moves will easily be accepted and the effects of the move on the total energy will not pose much of a constrain. If $kT$ is very low, on the other hand, moves that increase the total energy are very unlikely to be accepted and the system will likely be trapped in a local energy minimum instead of converging to an optimal global energy minimum. 

The definition of cells, movement and growth is rather simplistic in the CPM framework. While this may not be appropriate for all cell-based biological problems, the CPM framework has the great advantage of being relatively easy to implement. It avoids many computational problems of more sophisticated cell-based models, e.g. the boundaries of cells are clearly defined and cells cannot overlap due to the lattice structure.\\

\subsubsection{Agent-based Models}
Finally agent-based models can be used when cells take a more active role in moving in the tissue. It is then possible to consider the cells as interacting agents that move according to certain rules and that may serve as sources and sinks for extracellular proteins that then diffuse in the extracellular space. Time delays and non-linear responses can readily be incorporated. Agent-based cellular automata were originally introduced by John von Neumann and Stanislaw Ulam to study how complex biological behaviours might emerge from simple local rules. While agent-based models offer a great flexibility in encoding many details this comes at a heavy computational cost that limits the number of agents (cells) that can typically be followed. Agent-based models have been particularly popular in immunology where many behaviours depend on small cohorts of individual cells rather than tissues \cite{Bauer:2009wq}.
We have previously used agent-based models to model the germinal center reaction during an immune response with some 10000 cells \cite{MeyerHermann:2006cy}.
Parallel computing now permits the simulation of much larger systems and agent-based methods are also used in simulating morphogenic processes during development \cite{Thorne:2007kd}.


\section*{Acknowledgment}
%=================================================================================

The authors thank Erkan \"Unal, Javier Lopez-Rios und Dario Speziale from the Zeller lab for the embryo picture in Figure 1. The authors acknowledge funding from the SNF Sinergia grant "Developmental engineering of endochondral ossification from mesenchymal stem cells", a SystemsX RTD on Forebrain Development, a SystemsX iPhD grant, and an ETH Zurich postdoctoral fellowship to D.M..



\bibliographystyle{./IEEEtranBST/IEEEtran}
%\begin{thebibliography}{1}


%\bibitem{Horton2009}
%Horton, W. a, \& Degnin, C. R. (2009). \emph{FGFs in endochondral skeletal development},
%Trends in endocrinology and metabolism: TEM, 20(7), 341-8. doi:10.1016/j.tem.2009.04.003
%\end{thebibliography}

%\bibliography{/Users/iberd/MyDocuments/Academia/Publications/Bibliography/Library_Papers.bib}
%\bibliography{./bibliography/Library_Papers.bib}
%\bibliography{./bibliography/Library_tanakas.bib}
%\bibliography{./bibliography/Patrick.bib}

\bibliography{Library_papers}

\end{document}